\documentclass[sigconf,nonacm]{acmart}

\usepackage{booktabs}
\usepackage{tabularx}
\usepackage{array}
\usepackage{graphicx}
\usepackage{listings}
\usepackage{xcolor}
\usepackage{multirow}
\usepackage{enumitem}
\usepackage{pifont}
\usepackage{flafter}
\usepackage[ruled,lined,linesnumbered]{algorithm2e}
\graphicspath{{../figures/}{figures/}{./figures/}}

\definecolor{addGreen}{RGB}{117, 217, 125}

\newcommand{\hladd}[1]{%
\begingroup
\setlength{\fboxsep}{0pt}%
\colorbox{addGreen}{\ttfamily\scriptsize\strut #1}%
\endgroup
}

\newcommand{\finding}[2]{%
\begin{center}
\setlength{\fboxsep}{6pt}%
\colorbox{black!5}{%
\parbox{0.93\columnwidth}{\small\textbf{Finding #1.} #2}%
}%
\end{center}
}

\AtBeginDocument{%
  }
\AtBeginEnvironment{thebibliography}{\sloppy\raggedright\emergencystretch=8em\relax}

\title{DREA: Decoupled Reasoning and Exploration Agents for Repository-Level Vulnerability Detection}

\author{Mingyang Sun}
\email{sunmingyang@iie.ac.cn}
\affiliation{%
  \institution{Institute of Information Engineering, Chinese Academy of Sciences}
  \city{Beijing}
  \country{China}
}
\affiliation{%
  \institution{School of Cyber Security, University of Chinese Academy of Sciences}
  \city{Beijing}
  \country{China}
}

\author{Guozhu Meng}
\authornote{Corresponding author.}
\email{mengguozhu@iie.ac.cn}
\affiliation{%
  \institution{Institute of Information Engineering, Chinese Academy of Sciences}
  \city{Beijing}
  \country{China}
}
\affiliation{%
  \institution{School of Cyber Security, University of Chinese Academy of Sciences}
  \city{Beijing}
  \country{China}
}

\renewcommand{\shortauthors}{Sun and Meng}

\begin{document}

\begin{abstract}
Large language models (LLMs) are increasingly applied to vulnerability detection due to their strong code comprehension capabilities, but most existing approaches rely on isolated functions or context extracted by fixed program-analysis rules. These methods cannot adaptively explore repository-level dependencies to gather sufficient context when vulnerabilities span multiple functions or files, compromising detection reliability. 
We present DREA (Decoupled Reasoning and Exploration Agents), a hypothesis-driven framework for repository-level vulnerability detection. DREA decouples reasoning from exploration through two collaborating agents: a planning agent backed by an advanced LLM that forms vulnerability hypotheses and directs the investigation, and an explorer agent powered by a lightweight model that retrieves repository-level context on demand. Goal-directed context acquisition is the primary source of detection improvement in this design, while offloading token-heavy exploration to the local model keeps inference economically tractable.
To support evaluation, we construct RepoPairBench, a repository-grounded benchmark of validated Python vulnerability-fix pairs from real-world projects. Beyond binary detection accuracy, we introduce a reasoning correctness evaluation to assess whether a model's rationale matches the documented vulnerability mechanism.
Across three LLMs, DREA improves Pair-Correctness from $19\text{--}26\%$ to $30\text{--}42\%$ while offloading over 93\% of tokens to the explorer, reducing estimated billable API cost by a factor of 16--48. Reasoning correctness analysis further reveals that $26\text{--}55\%$ of true positives, for both DREA and the function-only baseline, are correct predictions supported by flawed rationales, identifying security reasoning quality as a shared bottleneck for current LLMs.
\end{abstract}

\keywords{vulnerability detection, LLM agents, repository-level analysis, reasoning evaluation, software security}

\ccsdesc[500]{Security and privacy~Software security engineering}
\ccsdesc[300]{Software and its engineering~Software creation and management}

\maketitle

\section{Introduction}

LLMs are increasingly applied to vulnerability detection~\cite{shengLLMsSoftwareSecurity2025,zhouLargeLanguageModel2024,kaniewskiSystematicLiteratureReview2025}, yet most existing approaches analyze isolated functions, optionally augmented with context extracted by fixed retrieval heuristics such as caller/callee retrieval~\cite{wenVulEvalRepositoryLevelEvaluation2024,wangReposVulRepositoryLevelHighQuality2024} or dependency-based snippet selection~\cite{yangContextEnhancedVulnerabilityDetection2025,lekssaysLLMxCPGContextAwareVulnerability2025}. Real-world vulnerabilities, however, often depend on repository-level evidence---cross-function data flows, sanitization or authorization logic implemented elsewhere, and project-specific configurations~\cite{risseTopScoreWrong2025}---and when such evidence lies outside the scope of pre-defined heuristics, these methods cannot adaptively gather the information needed for correct security judgments.

Practical vulnerability auditing is inherently hypothesis-driven: a human auditor forms tentative hypotheses about risky behavior and then selectively traces data flows, validation logic, and access-control checks to confirm or refute those hypotheses. Existing augmentation strategies, by contrast, apply the same retrieval heuristic regardless of the security hypothesis under investigation~\cite{duVulRAGEnhancingLLMbased2025,sunLLM4VulnUnifiedEvaluation2025,zhengLearningFocusContext2025}. This suggests a natural question: can LLM-based vulnerability detection benefit from \emph{hypothesis-driven repository exploration} rather than static context augmentation?

Recent LLM agents for software engineering provide a promising starting point. Systems such as SWE-Agent~\cite{yangSWEAgentAgentComputer2024}, Agentless~\cite{xiaAgentlessDemystifyingLLMbased2024}, and AutoCodeRover~\cite{zhangAutoCodeRoverAutonomousProgram2024} show that language-model-based agents can navigate repositories, inspect files, and complete complex engineering tasks such as bug fixing and issue resolution. Motivated by these advances, we investigate whether a similar paradigm can improve vulnerability detection: instead of classifying a target function using only pre-specified context, an agent could actively explore the repository, gather security-relevant evidence on demand, and use that evidence to support its decision.

\begin{figure*}[t]
\centering
\begin{minipage}[t]{0.48\textwidth}
\begin{lstlisting}[language=Python,basicstyle=\scriptsize\ttfamily]
def publish(self, id_, identity, uow=None):
    self.require_permission(identity, "publish")
    draft = self.draft_cls.pid.resolve(id_, registered_only=False)
    self._validate_draft(identity, draft)
    record = self.record_cls.publish(draft)
    self.run_components("publish", identity, draft=draft, record=record, uow=uow)
    uow.register(RecordCommitOp(record, indexer=self.indexer))
    uow.register(RecordDeleteOp(draft, force=False, indexer=self.indexer))
    return self.result_item(self, identity, record, links_tpl=self.links_item_tpl)
\end{lstlisting}
\centering\small (a) Vulnerable version.
\end{minipage}\hfill
\begin{minipage}[t]{0.48\textwidth}
\begin{lstlisting}[language=Python,basicstyle=\scriptsize\ttfamily,escapeinside={(*@}{@*)}]
def publish(self, id_, identity, uow=None):
    draft = self.draft_cls.pid.resolve(id_, registered_only=False)
(*@\hladd{    self.require\_permission(identity, "publish", record=draft)}@*)
    self._validate_draft(identity, draft)
    record = self.record_cls.publish(draft)
    self.run_components("publish", identity, draft=draft, record=record, uow=uow)
    uow.register(RecordCommitOp(record, indexer=self.indexer))
    uow.register(RecordDeleteOp(draft, force=False, indexer=self.indexer))
    return self.result_item(self, identity, record, links_tpl=self.links_item_tpl)
\end{lstlisting}
\centering\small (b) Patched version.
\end{minipage}
\caption{CVE-2021-43781 motivating example. The vulnerable \texttt{publish} method performs a global permission check, while the patch binds authorization to the resolved draft object via \texttt{record=draft}.}
\Description{Side-by-side code comparison of the vulnerable and patched publish method, with the added record-scoped authorization argument highlighted.}
\label{fig:motivation-example}
\end{figure*}

In this paper, we present \textbf{DREA} (\textbf{D}ecoupled \textbf{R}easoning and \textbf{E}xploration \textbf{A}gents), a hypothesis-driven framework for repository-level vulnerability detection.
\footnote{Our code and the RepoPairBench dataset are publicly available at \url{https://github.com/huhusmang/DREA}.}
 DREA enables a \emph{planning agent}, backed by a stronger reasoning model, to form vulnerability hypotheses, direct the investigation, and actively acquire repository-level evidence as hypotheses evolve---the primary source of its detection improvement over static approaches. To make this repository-scale exploration economically practical, DREA offloads the token-heavy navigation work to a lightweight \emph{exploration agent} deployed locally, keeping billable API cost tractable. To rigorously evaluate this framework, we additionally contribute a purpose-built benchmark and a complementary diagnostic evaluation protocol.

To evaluate repository-level vulnerability detection, we construct \textbf{RepoPairBench}, a benchmark of 100 validated Python vulnerability-fix pairs sourced from CVE records and linked to their complete Git repositories. Unlike widely used vulnerability datasets~\cite{zhouDevignEffectiveVulnerability2019,fanCCCodeVulnerability2020,dingVulnerabilityDetectionCode2024} that often emphasize C/C++ code or weaker label reliability, RepoPairBench targets recent Python vulnerabilities with pair-based evaluation and stricter repository-grounded validation. Details of the benchmark design are presented in Section~\ref{sec:benchmark}.

Beyond binary detection accuracy, we further introduce a \emph{reasoning correctness evaluation} that assesses whether a model's vulnerability rationale aligns with the documented vulnerability mechanism, using a structured LLM-as-a-Judge procedure. This diagnostic reveals cases in which a model reaches the correct label through flawed reasoning---a phenomenon we term \textbf{Lucky Hits}.

The contributions of this paper are as follows:

\begin{enumerate}
\item \textbf{A hypothesis-driven framework for repository-level vulnerability detection.} We propose DREA, which decouples security reasoning from repository exploration through collaborating Planner and Explorer agents, enabling goal-directed context acquisition while keeping analysis cost tractable.

\item \textbf{A repository-grounded Python benchmark for vulnerability detection.} To support this evaluation, we construct RepoPairBench, a benchmark of 100 carefully validated CVE-sourced Python vulnerability-fix pairs from 2021--2025, spanning 48 CWE categories.

\item \textbf{A reasoning correctness evaluation protocol for vulnerability detection.} We introduce a structured LLM-as-a-Judge procedure that assesses whether a model's vulnerability rationale aligns with the documented vulnerability mechanism, enabling diagnosis of correct predictions supported by flawed reasoning (Lucky Hits).

\item \textbf{A comprehensive empirical study.} Across three backbone models and four detection configurations, we show that DREA improves Pair-Correctness from 19--26\% to 30--42\% while reducing estimated billable API cost by a factor of 16--48. Ablation experiments attribute this gain to DREA's structured exploration rather than context volume alone, and reasoning correctness analysis reveals that 26--55\% of true positives are Lucky Hits, identifying security reasoning quality as a shared bottleneck for current LLMs.
\end{enumerate}

\begin{figure*}[t]
\centering
\includegraphics[width=0.92\textwidth,height=0.72\textheight,keepaspectratio]{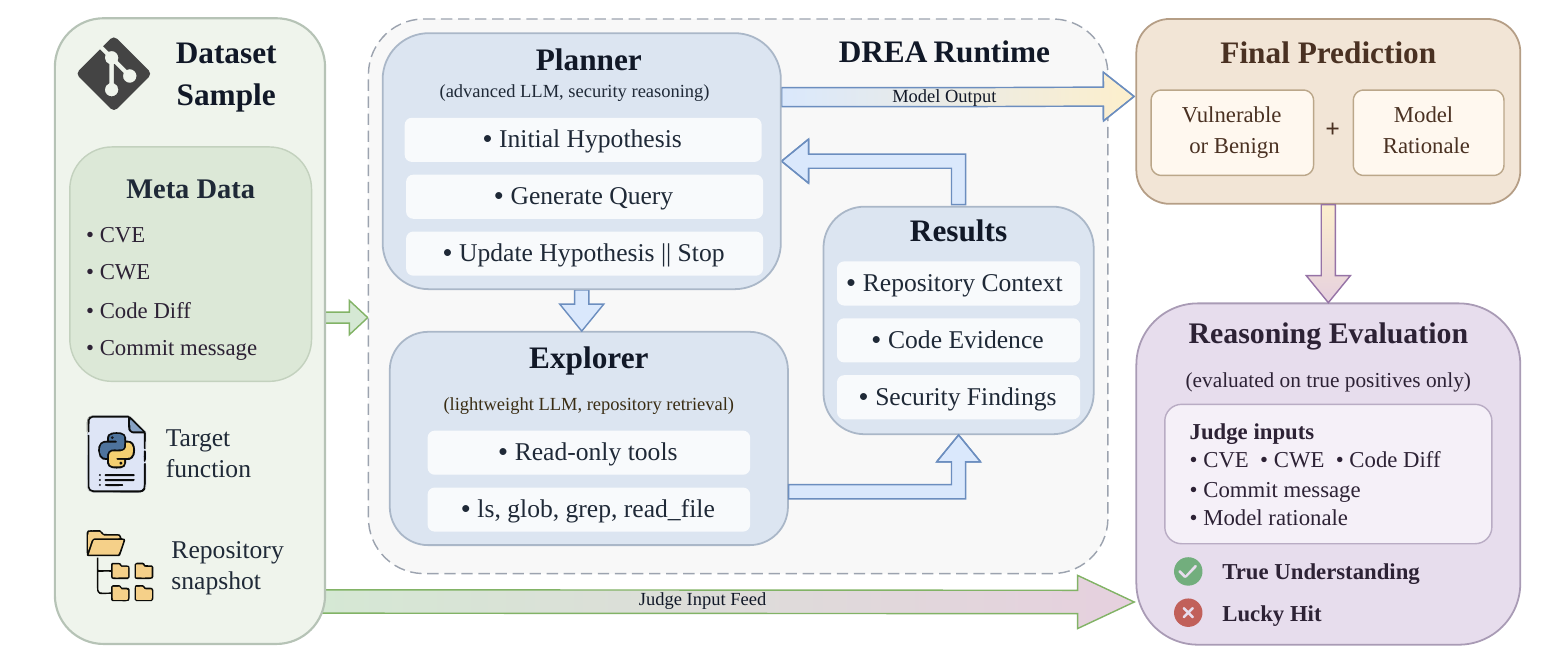}
\caption{DREA architecture. The Planner revises security hypotheses and delegates evidence-gathering requests to a local Explorer, which navigates the repository with read-only tools and returns structured findings.}
\Description{Architecture diagram of DREA showing the Planner--Explorer interaction loop with repository navigation tools.}
\label{fig:1}
\end{figure*}

\section{Motivating Example}
We illustrate the challenge addressed by DREA with a real-world case from \texttt{invenio-drafts-resources} (CVE-2021-43781, fix commit \href{https://github.com/inveniosoftware/invenio-drafts-resources/commit/039b0cff1ad4b952000f4d8c3a93f347108b6626}{\texttt{039b0cff}}). The vulnerability cannot be correctly identified from the target function alone, because the decisive evidence lies in repository-level permission logic that requires cross-file exploration.

\paragraph{Why function-level analysis fails.}
Figure~\ref{fig:motivation-example} shows CVE-2021-43781 from \texttt{invenio-drafts-resources}. The vulnerable \texttt{publish} method appears protected because it calls \texttt{require\_permission(identity, "publish")} before publishing. A function-level analyzer treats this call as proof that the workflow is guarded, but from the function alone the missing detail is invisible: the check is not bound to the draft selected by \texttt{id\_}, so it verifies only a global capability rather than ownership of the concrete object.

\paragraph{How DREA recovers the trigger path.}
DREA recovers the trigger path by collecting three cross-file facts: (1)~sibling methods such as \texttt{read\_draft}, \texttt{edit}, and \texttt{delete\_draft} all pass \texttt{record=draft} for object-scoped authorization; (2)~the default \texttt{can\_publish} policy is \texttt{[AnyUser()]}, meaning the global check is maximally permissive; and (3)~the target call omits the \texttt{record} argument. The patch (Figure~\ref{fig:motivation-example}b) confirms the root cause: it resolves the draft first and then calls \texttt{require\_permission(..., record=draft)}, binding authorization to the concrete object. Together, these facts expose the trigger path: an attacker-controlled \texttt{id\_} selects a draft, while the permission check validates only a global capability and never binds the authorization decision to that draft. This case exemplifies a broader pattern: verifying the \emph{presence} of a security mechanism is insufficient without verifying its \emph{scope}---a distinction that requires cross-file evidence. We formalize this approach in the next section.

\section{Methodology}

This section presents our methodology. We first present \textbf{DREA}, our hypothesis-driven framework for repository-level vulnerability detection (Section~\ref{sec:drea}). We then describe \textbf{RepoPairBench}, a repository-grounded benchmark constructed to evaluate this setting (Section~\ref{sec:benchmark}). Finally, we introduce a complementary \textbf{reasoning correctness evaluation} that provides diagnostic insight beyond standard binary metrics (Section~\ref{sec:eval-protocol}).

\subsection{DREA: Decoupled Reasoning and Exploration Agents}
\label{sec:drea}

The key idea of DREA is hypothesis-driven context acquisition: instead of appending a fixed amount of surrounding code to every sample, DREA acquires repository evidence on demand and updates its exploration targets as the vulnerability hypothesis evolves. This active, goal-directed exploration is the primary source of DREA's detection improvement over static approaches.

Because repository-level analysis requires extensive code navigation---tracing data flows across files, locating validation logic, and inspecting callers---the resulting token volume would be prohibitively expensive if routed entirely through a high-cost API model. DREA therefore decouples the workflow into two collaborating agents, shifting the token-heavy exploration cost from the expensive reasoning model to a lightweight locally deployed model. DREA consists of two agents:

\begin{itemize}
    \item \textbf{Planner}: a stronger reasoning model responsible for security reasoning, planning, and the final vulnerability decision.
    \item \textbf{Explorer}: a lightweight locally deployed model responsible for repository navigation and evidence retrieval.
\end{itemize}

This division of labor keeps the Planner's context focused on security semantics while shifting the token-heavy repository traversal to the Explorer, making the design both cost-efficient and scalable to large repositories. Figure~\ref{fig:1} shows the overall architecture.

\paragraph{Planner.}
Starting from the target function, the Planner identifies security-relevant operations, forms an initial vulnerability hypothesis, and determines what repository evidence is needed next. It then delegates goal-directed exploration requests to the Explorer, such as tracing callers, locating validation logic, or finding authorization checks. After receiving the Explorer's findings, it updates the hypothesis and either continues exploration or issues a final decision. In our prompting setup, a \texttt{VULNERABLE} output must describe a plausible trigger path from attacker-controlled input to a dangerous operation or violated security boundary, whereas a \texttt{BENIGN} output must argue that the relevant operations are adequately guarded. Algorithm~\ref{alg:drea} formalizes this iterative protocol.

\paragraph{Explorer.}
The Explorer operates with four read-only tools---\texttt{ls}, \texttt{glob}, \texttt{grep}, and \texttt{read\_file}---and is restricted to gathering repository facts rather than making vulnerability judgments. Given a Planner request, it navigates the repository and returns structured findings organized into three categories:
\begin{itemize}[leftmargin=1.4em,itemsep=0.15em,topsep=0.2em,parsep=0pt]
\item \textbf{Repository Context}: background information that situates the target function within the broader repository, such as relevant files, call relationships, surrounding modules, data structures, and execution flow. This helps the Planner determine where to investigate next.
\item \textbf{Code Evidence}: concrete implementation details including matched code patterns, key statements, and relevant code excerpts. This provides the direct factual basis that the Planner uses to verify or refute its current vulnerability hypothesis.
\item \textbf{Security Findings}: security-relevant observations synthesized from the retrieved repository facts, such as missing checks, inconsistent authorization patterns, suspicious data-flow links, or other repository-level cues that may affect the vulnerability assessment. These do not constitute the final judgment; rather, they help the Planner refine its reasoning and decide whether further exploration is needed.
\end{itemize}
Because the Explorer's task is retrieval rather than reasoning, it can be implemented using a lightweight, locally deployed model, significantly reducing analysis cost (implementation details in Section~\ref{sec:setup}).

\begin{algorithm}[t]
\caption{DREA Interaction Protocol}
\label{alg:drea}
\KwIn{Target function $x$, repository $\mathcal{R}$, interaction budget $B$}
\KwOut{Vulnerability decision $d \in \{\textsc{Vulnerable}, \textsc{Benign}\}$}

\textbf{Step 1: Initialize Hypothesis}\;
Planner inspects $x$, identifies security-relevant operations, and forms initial vulnerability hypothesis $h_0$\;
$h \leftarrow h_0$\;

\textbf{Step 2: Iterative Exploration}\;
\For{$t = 1$ \KwTo $B$}{
    Planner generates exploration requests $Q_t$ based on current hypothesis $h$\;
    Explorer executes $Q_t$ using read-only tools (\texttt{ls}, \texttt{glob}, \texttt{grep}, \texttt{read\_file}) over $\mathcal{R}$\;
    Explorer returns structured findings $F_t$ (repository context, code evidence, security findings)\;
    Planner incorporates $F_t$ and updates hypothesis: $h \leftarrow \textsc{Update}(h, F_t)$\;
    \If{Planner determines evidence is sufficient}{
        \textbf{break}\;
    }
}

\textbf{Step 3: Final Decision}\;
\eIf{$h$ describes a plausible trigger path from attacker-controlled input to a dangerous operation}{
    \Return $d = \textsc{Vulnerable}$ with trigger path description\;
}{
    \Return $d = \textsc{Benign}$ with guarding evidence\;
}
\end{algorithm}

A typical analysis in our experiments involves approximately 10 Planner--Explorer rounds per sample. This iterative protocol concentrates expensive model capacity on security reasoning while shifting token-heavy search steps to the local Explorer.

\subsection{RepoPairBench: A Repository-Grounded Benchmark for Python Vulnerability Pairs}
\label{sec:benchmark}

Evaluating repository-level vulnerability detection requires more than isolated code snippets: the benchmark must preserve the relation between a vulnerable function, its patch, and the repository context needed to interpret the vulnerability. Following the spirit of PrimeVul~\cite{dingVulnerabilityDetectionCode2024}, we retain paired vulnerable/fixed instances, but target a different setting: recent, repository-grounded Python vulnerabilities for which systems may inspect the full project rather than classify a function in isolation.

\paragraph{Data collection and filtering.}
We collect CVE-linked fixing commits from the National Vulnerability Database (NVD) and extract function-level diffs using PyDriller~\cite{spadiniPyDrillerPythonFramework2018}. From these candidates, we construct \textbf{RepoPairBench}, a benchmark of 100 vulnerability-fix pairs for Python projects. Each pair contains (1) the vulnerable function from the parent of the fixing commit, (2) the patched version after the fix, and (3) the repository URL and commit hash needed to reconstruct the full project state. Each pair is further annotated with its CVE identifier(s), Common Weakness Enumeration (CWE) category label(s), and the fixing commit message, which serve as ground truth for the reasoning evaluation described in Section~\ref{sec:eval-protocol}. To improve attribution quality, we apply strict filtering: the function must align cleanly before and after the fix, and the commit must be \emph{modification-only}, excluding file additions, deletions, and newly introduced functions. Patched instances are labeled \texttt{BENIGN} with respect to the specific documented vulnerability; this label does not imply the absence of other, unrelated security issues. If a model flags a patched function for a different vulnerability mechanism, the prediction is counted as a false positive under our paired protocol.

\paragraph{Scope and positioning.}
RepoPairBench focuses on recent Python vulnerabilities disclosed between 2021 and 2025. We choose this scope for three reasons: (1)~recent samples better reflect contemporary libraries and coding practices, reducing label noise from outdated patterns~\cite{gaoMonoYourClean2025,liCleanVulAutomaticFunctionLevel2025}; (2)~Python is among the most actively maintained interpreted languages on GitHub with a substantial volume of CVE-linked fixing commits, providing a sufficiently large pool for rigorous filtering; and (3)~Python's dynamic typing and extensive use in web frameworks (Django, Flask) and data pipelines expose it to a diverse range of vulnerability types---injection, deserialization, path traversal---that naturally require cross-file context to diagnose. The benchmark contains 100 vulnerability-fix pairs (200 individual instances) spanning 48 CWE categories; the five most frequent are CWE-79 (13), CWE-22 (10), CWE-20 (7), CWE-502 (7), and CWE-94 (7).

\paragraph{Repository-grounded evaluation setup.}
For each pair, we prepare two repository snapshots: the \emph{parent commit} of the fixing commit, representing the vulnerable project state, and the \emph{fixing commit} itself, representing the patched state. The vulnerable function is analyzed against the pre-fix snapshot and the patched function against the post-fix snapshot, ensuring that each instance is evaluated with a repository context consistent with its code version. The agent receives the target function and its file path, and may inspect the checked-out repository using read-only navigation tools. This setup mirrors the practical scenario in which an auditor starts from a suspicious function and investigates surrounding project context on demand.

\subsection{Reasoning Correctness Evaluation}
\label{sec:eval-protocol}
To provide diagnostic insight beyond standard binary metrics, we additionally evaluate whether a model's vulnerability rationale aligns with the documented vulnerability mechanism. We assess outputs along two dimensions: \emph{conclusion correctness}, which is computed on all vulnerable and patched instances, and \emph{reasoning quality}, which measures whether the explanation aligns with the documented vulnerability mechanism. We assess reasoning quality only on \emph{true positive} vulnerable samples, where the model predicts \texttt{VULNERABLE} and its explanation can be compared meaningfully against the CVE and the corresponding fix.

\paragraph{Judge inputs and criteria.}
We operationalize reasoning quality using an LLM-as-a-Judge procedure~\cite{zhengJudgingLLMasaJudge2024}. We select GPT-4.1 as the judge model for two reasons: (1)~it offers strong semantic understanding and instruction-following capability, which are essential for faithfully applying our judging rubric to long, heterogeneous inputs (CVE text, diffs, and model rationales), and (2)~it is not used as a backbone in any of our evaluated configurations (GPT-5.2, DeepSeek-V3.2, GLM-4.7), thereby avoiding self-evaluation bias in which a model favors its own reasoning patterns~\cite{zhengJudgingLLMasaJudge2024}. For each true positive vulnerable sample, the judge receives (1) the CVE description, (2) the fix commit message, (3) the code diff between the vulnerable and patched versions, (4) the CWE labels, and (5) the model's final analysis output (the last assistant-generated text from the planner agent). The judge then determines whether the explanation is consistent with the documented vulnerability. Our judging protocol evaluates four criteria:
\begin{enumerate}
    \item whether the identified vulnerability type aligns with the documented CWE category;
    \item whether the analysis correctly pinpoints the relevant vulnerable code elements;
    \item whether the analysis correctly identifies the root cause of the vulnerability;
    \item whether the described exploitation mechanism is consistent with the actual vulnerability.
\end{enumerate}

The judge outputs a structured JSON record, enforced through a Pydantic schema, containing both the assessment outcome and an explanatory critique. Our evaluation is intentionally strict: if the system identifies a plausible but different vulnerability from the one documented by the CVE and fix, we count the reasoning as inaccurate.

\paragraph{Lucky Hits.}
Combining the two dimensions reveals a class of cases that binary metrics cannot distinguish: predictions with correct labels but incorrect reasoning. We refer to such cases as \textbf{Lucky Hits}. Formally, among true positives,
\[
\mathrm{LuckyHitRate}
=
\frac{|\{x \in TP : \mathrm{Reasoning}(x)=\mathrm{Inaccurate}\}|}{|TP|}.
\]
A high Lucky Hit Rate (LHR) indicates that many apparent successes under binary evaluation do not correspond to sound vulnerability reasoning. Reasoning Accuracy (RA) is defined symmetrically on the same denominator: $\mathrm{RA} = 1 - \mathrm{LHR}$, i.e., the fraction of true positives with accurate reasoning.

\section{Experimental Setup}
\label{sec:setup}

\subsection{Task, Models, and Baseline}

We evaluate on RepoPairBench (Section~\ref{sec:benchmark}), which yields 200 classification instances (100 vulnerable, 100 patched) each grounded in its corresponding repository snapshot.
We evaluate three LLMs as Planner backbones: \textbf{GPT-5.2}, \textbf{DeepSeek-V3.2}, and \textbf{GLM-4.7}. For the Explorer, we use \textbf{GLM-4.7-Flash} (4-bit AWQ quantization) deployed locally via vLLM on a single A800 GPU, supporting 256K context inference across all DREA runs. All models use default decoding parameters (temperature and top-$p$). Fixing the Explorer across conditions reduces variation in the retrieval component and allows differences across configurations to more directly reflect Planner behavior.

To isolate the contributions of different factors, we compare DREA against three controlled baselines. The \textbf{Function-Only Baseline} uses the same backbone model and output schema but receives only the target function, with no repository access or tool use. The \textbf{Whole-File Baseline} additionally provides the complete source file containing the target function but without exploration tools. The \textbf{Single-Agent Baseline} gives the same backbone model direct access to the Explorer's read-only tools without the Planner--Explorer separation. Ablation results for the latter two baselines are reported in Table~\ref{tab:ablation}.

\subsection{Evaluation Metrics}
We evaluate model performance along four complementary dimensions:
\begin{enumerate}
    \item \textbf{Standard detection metrics.}
    Recall, False Positive Rate (FPR), and F\textsubscript{1}, computed over all 200 instances.
    \item \textbf{Pair-based metrics.}
    We evaluate the 100 vulnerability--fix pairs following the paired evaluation protocol of PrimeVul~\cite{dingVulnerabilityDetectionCode2024}. Our primary pair-level metric is Pair-Correctness (P-C), which requires \emph{both} members of a pair to be classified correctly. We additionally report the pair-level outcome decomposition used later in Table~\ref{tab:2}: P-V (both predicted vulnerable), P-B (both predicted benign), and P-R (labels reversed). These auxiliary pair-level categories are not headline metrics, but they help explain how errors shift between methods.
    \item \textbf{Reasoning metrics.}
    RA and LHR (defined in Section~\ref{sec:eval-protocol}), both computed over the true-positive subset.
    \item \textbf{Overall discrimination.}
    Youden's $J$ statistic ($J = \mathrm{Recall} - \mathrm{FPR}$).
\end{enumerate}

\subsection{Judge Reliability}
\label{subsec:judge-reliability}
Because reasoning-quality assessment relies on an LLM-as-a-Judge component, we validate the judge against independent human annotations. We sample 50 judged cases using stratified random sampling over model and method conditions. Two security researchers with more than five years of vulnerability-analysis experience independently annotate the sampled cases. Each annotator receives the same evidence available to the automated judge: the ground-truth context (including the CVE description, fix commit message, code diff, and CWE label) together with the model's final analysis output. We then compare each human annotator against the LLM judge using raw agreement and Cohen's $\kappa$, and also report inter-annotator agreement. 

Agreement between the LLM judge and the two annotators is 96.0\% and 94.0\%, corresponding to Cohen's $\kappa$ values of 0.92 and 0.88, respectively. Inter-annotator agreement is 92.0\% with $\kappa = 0.84$. These results indicate that the automated judge is sufficiently reliable for the reasoning-quality analysis reported in this paper.

\section{Results and Analysis}
We organize the empirical analysis around five research questions. The overall picture is that DREA consistently improves detection reliability over function-level analysis while remaining cost-efficient through the Planner--Explorer separation. We additionally examine reasoning correctness and find that flawed rationales behind correct predictions are a shared bottleneck across all evaluated LLMs. The remainder of this section addresses: (RQ1) Does the agent-based approach improve detection reliability? (RQ2) What is the cost structure of the dual-agent design? (RQ3) How do detection capabilities vary across vulnerability types and models? (RQ4) Does more exploration lead to better analysis? (RQ5) What does reasoning correctness evaluation reveal about LLM security reasoning?
\subsection{RQ1: Does Agent-Based Detection Improve Reliability?}
In this RQ, we evaluate whether repository-level agentic exploration improves detection reliability relative to a controlled Function-Only Baseline. Table~\ref{tab:1} reports the main results, using P-C and Youden's $J$ as our primary indicators of pair-level reliability and balanced discrimination.

\begin{table}[htbp]
\centering
\footnotesize
\caption{Detection performance of DREA versus the Function-Only Baseline on RepoPairBench. P-C denotes Pair-Correctness. The best P-C and Youden's $J$ for each backbone are \textbf{bolded}.}
\label{tab:1}
\resizebox{\columnwidth}{!}{%
\begin{tabular}{@{}ll rrrrr@{}}
\toprule
Model & Method & Recall (\%) & FPR (\%) & $F_1$ (\%) & P-C (\%) & Youden's $J$ (pp)\\
\midrule
\multirow{2}{*}{DeepSeek-V3.2} & DREA & 80.0 & 45.0 & 71.1 & \textbf{42.0} & \textbf{35.0}\\
 & Function-Only & 39.0 & 32.0 & 45.6 & 19.0 & 7.0\\
\midrule
\multirow{2}{*}{GLM-4.7} & DREA & 59.0 & 38.0 & 59.9 & \textbf{34.0} & \textbf{21.0}\\
 & Function-Only & 54.0 & 43.0 & 54.8 & 26.0 & 11.0\\
\midrule
\multirow{2}{*}{GPT-5.2} & DREA & 53.0 & 28.0 & 58.6 & \textbf{30.0} & \textbf{25.0}\\
 & Function-Only & 42.0 & 25.0 & 50.3 & 21.0 & 17.0\\
\bottomrule
\end{tabular}
}
\end{table}

The pattern is consistent across all three backbones: DREA achieves the best P-C in every case, improving DeepSeek-V3.2 from 19\% to 42\%, GLM-4.7 from 26\% to 34\%, and GPT-5.2 from 21\% to 30\%. Youden's $J$ is likewise highest for every DREA configuration, indicating that the gains are not due to a trivial recall increase accompanied by excessive false positives. Because the baseline uses the same backbone and output format, this comparison isolates repository access and agentic exploration as the principal difference between the two settings.

A closer look at the recall-FPR dynamics reveals important differences in how each backbone responds to repository access. DeepSeek-V3.2 exhibits the most aggressive shift: recall rises by 41 percentage points while FPR increases by only 13, yielding the largest Youden's $J$ improvement (+28 points). GLM-4.7 \emph{reduces} its FPR from 43\% to 38\% while increasing recall, indicating that repository context can suppress false alarms rather than simply increasing detection aggressiveness. GPT-5.2 is the most conservative, adding 11 points of recall at the cost of only 3 points of FPR. Across all backbones, recall gains substantially exceed FPR increases, confirming that repository exploration converts more false negatives into true positives than it converts true negatives into false positives.

Table~\ref{tab:2} helps explain where these gains come from. The Function-Only Baseline shows consistently high P-B counts (DeepSeek-V3.2: 49, GLM-4.7: 31, GPT-5.2: 54), indicating a conservative tendency to default to \texttt{BENIGN} when the local function does not provide enough evidence. DREA reduces these conservative misses and converts a substantial fraction of them into P-C outcomes, while leaving P-R counts roughly stable. In other words, repository exploration improves reliability primarily by resolving uncertainty in borderline cases, not by indiscriminately over-reporting vulnerabilities.

\begin{table}[htbp]
\centering
\footnotesize
\caption{Pair-level outcome decomposition on RepoPairBench. For each model--method setting, values are counts over 100 pairs, with each pair belonging to exactly one category: P-C, P-V, P-B, or P-R. The best P-C for each backbone is \textbf{bolded}.}
\label{tab:2}
\setlength{\tabcolsep}{6.5pt}
\begin{tabular*}{\columnwidth}{@{\extracolsep{\fill}}llrrrr@{}}
\toprule
Model & Method & P-C & P-V & P-B & P-R\\
\midrule
\multirow{2}{*}{DeepSeek-V3.2} & DREA & \textbf{42} & 38 & 13 & 7\\
 & Function-Only & 19 & 20 & 49 & 12\\
 \midrule
\multirow{2}{*}{GLM-4.7} & DREA & \textbf{34} & 25 & 28 & 13\\
 & Function-Only & 26 & 28 & 31 & 15\\
\midrule
\multirow{2}{*}{GPT-5.2} & DREA & \textbf{30} & 23 & 42 & 5\\
 & Function-Only & 21 & 21 & 54 & 4\\
\bottomrule
\end{tabular*}%
\end{table}

The P-V column reveals a complementary pattern: for DeepSeek-V3.2, P-V nearly doubles from 20 to 38, reflecting DREA's higher recall catching the vulnerable member while moderately elevated FPR causes the patched member to also be flagged. P-R counts remain stable across all backbones (12$\to$7, 15$\to$13, 4$\to$5). Taken together, DREA's principal benefit is converting conservative misses into correct pair-level detections; the P-V increase is a byproduct of the same recall--FPR trade-off.

\paragraph{Ablation: isolating the source of improvement.}
\label{sec:ablation}

To isolate the source of DREA's improvement, we compare it against the Whole-File and Single-Agent baselines defined in Section~\ref{sec:setup}. We conduct this ablation on DeepSeek-V3.2, the backbone with the largest DREA improvement, and restrict it to a single backbone because each agentic run consumes over one million tokens per sample (Section~\ref{sec:rq2-cost}), making a full cross-model sweep computationally prohibitive.

\begin{table}[htbp]
\centering
\footnotesize
\caption{Ablation study on DeepSeek-V3.2. The best P-C is \textbf{bolded}. API Tokens denotes the approximate mean per-sample token count routed through the paid API, rounded to the nearest thousand.}
\label{tab:ablation}
\resizebox{\columnwidth}{!}{%
\begin{tabular}{@{}l rrrr r@{}}
\toprule
Method & Recall (\%) & FPR (\%) & $F_1$ (\%) & P-C (\%) & API Tokens (K)\\
\midrule
DREA & 80.0 & 45.0 & 71.1 & \textbf{42.0} & 88\\
Single-Agent & 73.0 & 64.0 & 61.6 & 24.0 & 442\\
Whole-File & 57.0 & 41.0 & 57.6 & 26.0 & 20\\
Function-Only & 39.0 & 32.0 & 45.6 & 19.0 & 3\\
\bottomrule
\end{tabular}
}
\end{table}

Table~\ref{tab:ablation} reveals a clear hierarchy: Function-Only $<$ Whole-File $<$ DREA, with Single-Agent performing \emph{worse} than Whole-File despite full exploration capability. Whole-File improves modestly over Function-Only (+7 percentage points in P-C) but stays far below DREA, with a rising FPR (32\%$\to$41\%) that signals undirected context introduces noise. Single-Agent reaches only 24\% P-C at 64\% FPR: it accumulates 442K tokens of raw code (vs.\ 88K for DREA's Planner), and this context overload likely drives over-reporting. Since Single-Agent and DREA share the same exploration capability and differ mainly in whether retrieved code is filtered into structured evidence, the Planner--Explorer separation appears to contribute beyond cost reduction by mitigating context overload.\footnote{As the Single-Agent baseline uses one model for both roles while DREA uses a separate Explorer model, this effect cannot be fully isolated from model-level differences.} These results support the conclusion that DREA's gain stems from structured exploration rather than context volume or tool access alone.

\finding{1}{Repository-level agentic exploration improves detection reliability over function-level analysis across all three evaluated backbones. Ablation further indicates that the gain comes from structured exploration rather than context volume or tool access alone.}

\subsection{RQ2: Cost-Efficiency of Planner-Explorer Separation}
\label{sec:rq2-cost}

In this RQ, we assess whether the Planner--Explorer separation makes repository-level analysis economically practical. Table~\ref{tab:5} shows that across all three models, the Planner accounts for only 2.1--6.3\% of total token consumption, while the Explorer handles the remaining 93.7--97.9\% locally.

\begin{table}[htbp]
\centering
\small
\caption{Average per-sample token consumption under the Planner--Explorer separation. API Token~\% indicates the fraction of total tokens routed through the paid Planner API. Cost Reduction estimates the reduction in billable API cost relative to a hypothetical single-model design where all tokens incur API cost.}
\label{tab:5}
\resizebox{\columnwidth}{!}{%
\begin{tabular}{@{}l rrr rr@{}}
\toprule
Model & Total Tokens & Planner & Explorer & API Token \% & Cost Reduction\\
\midrule
DeepSeek-V3.2 & 1,397,669 & 87,847 & 1,309,822 & \textbf{6.3\%} & $\sim$16$\times$\\
GLM-4.7 & 1,560,519 & 32,557 & 1,527,962 & \textbf{2.1\%} & $\sim$48$\times$\\
GPT-5.2 & 345,691 & 8,357 & 337,334 & \textbf{2.4\%} & $\sim$41$\times$\\
\bottomrule
\end{tabular}
}
\end{table}

In a hypothetical single-model design, the full 0.3--1.6M tokens per sample would be billed at API rates. Under the decoupled design, only 8K--88K Planner tokens incur API cost, reducing estimated billable API cost to 2.1\%--6.3\% of that under a hypothetical single-model design (a factor of 16--48$\times$). This estimate reflects billable API token savings; it does not account for local GPU inference cost of the Explorer, which we deploy on a single A800 GPU. The main value of decoupling is therefore not merely architectural neatness, but the ability to preserve repository-level analysis while charging only a small fraction of tokens to the expensive model.


\finding{2}{Separating Planner and Explorer makes repository-grounded analysis practical: 93.7--97.9\% of tokens are offloaded to the local model, reducing estimated billable API cost by a factor of 16--48.}

\subsection{RQ3: Fine-Grained Analysis}

\subsubsection{CWE-Wise Performance}
\begin{figure}[htbp]
\centering
\includegraphics[width=0.96\columnwidth,trim=0 0 0 0,clip]{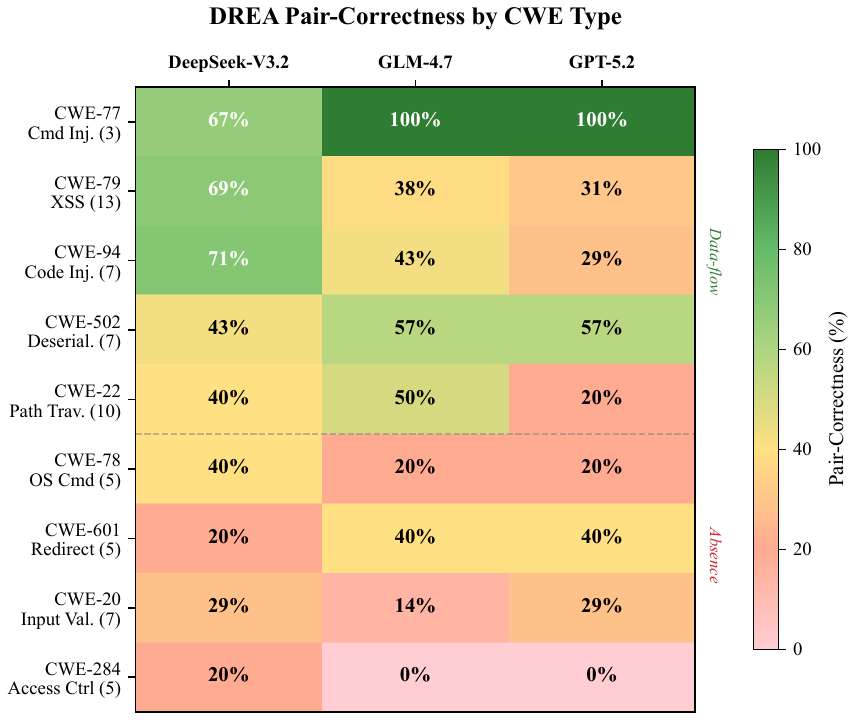}
\caption{Pair-Correctness (\%) of three DREA backbones stratified by CWE category. Categories are ordered by mean P-C: data-flow vulnerabilities (top rows) consistently outperform absence-type flaws (bottom rows). Parenthetical values indicate the number of pairs per category.}
\Description{Heatmap showing Pair-Correctness percentages for DeepSeek-V3.2, GLM-4.7, and GPT-5.2 across CWE categories, ordered from highest to lowest mean performance.}
\label{fig:4}
\end{figure}

In this RQ, we ask how DREA's behavior varies across vulnerability classes and model backbones, and what kinds of reasoning failures underlie these differences. We begin with heterogeneity by CWE type. Figure~\ref{fig:4} reveals a clear dichotomy: vulnerability types with explicit data-flow structure are substantially easier than those defined by the absence of a protection mechanism.

For example, DeepSeek-V3.2 reaches 69.2\% P-C on CWE-79 and 71.4\% on CWE-94, while CWE-502 reaches 42.9--57.1\% across models. In contrast, performance on CWE-284 falls to 0--20\%, and CWE-20 remains at 14.3--28.6\%. This contrast is intuitive. Source--sink vulnerabilities often provide observable syntactic and semantic cues, whereas missing-protection vulnerabilities require reasoning about what \emph{should} be present but is not. These per-CWE results should nevertheless be interpreted cautiously, since some subgroups are small.

\subsubsection{Model Behavioral Profiles}
The backbones also exhibit distinct operating regimes. DeepSeek-V3.2 is the most aggressive configuration: it attains the highest recall (80\%) and P-C (42\%), but also the highest FPR (45\%) and Lucky Hit Rate (55\%). GPT-5.2 lies at the opposite end, with the lowest recall (53\%) but also the lowest FPR (28\%) and Lucky Hit Rate (32.1\%). GLM-4.7 occupies an intermediate position. These profiles shift the operating point along a coverage--reliability trade-off: DeepSeek-V3.2 suits recall-oriented security audits where missing a vulnerability is costly, while GPT-5.2 suits analyst-time-constrained triage where each reported finding must be actionable.

Qualitative inspection of the judge critiques reveals a recurring pattern: when reasoning fails, models frequently default to generic ``insufficient input validation'' explanations regardless of the actual vulnerability type. This tendency is consistent with the CWE-level results above, where absence-type flaws (which cannot be explained by input validation alone) exhibit the lowest performance.

\finding{3}{On the evaluated benchmark, performance varies systematically by vulnerability structure and model operating regime: DREA shows the strongest results on explicit data-flow vulnerabilities and the weakest on absence-type flaws, with input-validation fixation emerging as the dominant reasoning error across backbones.}

\subsection{RQ4: Does More Exploration Lead to Better Analysis?}

In this RQ, we examine whether deeper exploration leads to better analysis. Table~\ref{tab:3} presents DeepSeek-V3.2 as a representative case, comparing token consumption between successful and failed trajectories.

\begin{table}[htbp]
\centering
\footnotesize
\caption{Mean per-sample token consumption and tool-call count by prediction outcome for the DeepSeek-V3.2 DREA configuration. Columns partition the 200 instances into true positives (TP), false negatives (FN), true negatives (TN), and false positives (FP); instance counts are shown in parentheses.}
\label{tab:3}
\resizebox{\columnwidth}{!}{%
\begin{tabular}{@{}l rrrr@{}}
\toprule
 & TP\,(80) & FN\,(20) & TN\,(55) & FP\,(45)\\
\midrule
Total tokens & 1,241,766 & \textbf{1,704,579} & 1,473,119 & 1,446,207\\
Explorer tokens & 1,153,822 & \textbf{1,625,495} & 1,386,132 & 1,353,586\\
Tool calls & 9.9 & 9.2 & 10.0 & 10.0\\
\bottomrule
\end{tabular}
}
\end{table}

Counterintuitively, incorrect vulnerable cases consume substantially \emph{more} exploration than correct ones: false negatives on vulnerable samples use 37\% more total tokens (1.7M vs.\ 1.2M) and 41\% more Explorer tokens than correct detections. Across all DREA configurations, the Pearson correlation between per-sample token consumption and binary prediction correctness is approximately $r=-0.34$, indicating a weak but consistent negative relationship.

This pattern suggests that higher exploration volume is often a symptom of analytical difficulty rather than a driver of better performance. When the vulnerability mechanism is identifiable from repository evidence, the Planner typically reaches the relevant findings quickly; when the backbone's reasoning capacity is insufficient, further exploration accumulates code without corresponding analytical leverage. The dominant bottleneck therefore lies in reasoning quality rather than information access---a finding consistent with the Lucky Hit analysis in RQ5.

The asymmetry between vulnerable and secure cases offers an additional insight. For vulnerable samples, failed cases consume 37\% more tokens than successful ones (1.70M vs.\ 1.24M); for secure samples, the difference is negligible (1.45M vs.\ 1.47M). Detecting a vulnerability requires locating a specific exploitable mechanism, which the Planner can achieve efficiently when the evidence is clear; confirming security requires systematically ruling out threat hypotheses. The mean number of tool calls is nearly identical across all outcome categories (9.2--10.0), indicating that the bottleneck lies in how effectively the model synthesizes results rather than how many queries it makes.

\finding{4}{More exploration does not reliably yield better analysis. Failed trajectories are often longer and more expensive than successful ones, indicating that the dominant bottleneck lies in reasoning rather than information access.}

\subsection{RQ5: What Does Reasoning Correctness Evaluation Reveal?}
In this RQ, we apply the reasoning correctness evaluation introduced in Section~\ref{sec:eval-protocol} to examine whether correct detection labels are supported by sound vulnerability rationales. This analysis applies to both DREA and the Function-Only Baseline, revealing a shared pattern across all evaluated LLMs.

\begin{figure}[htbp]
\centering
\includegraphics[width=0.96\columnwidth]{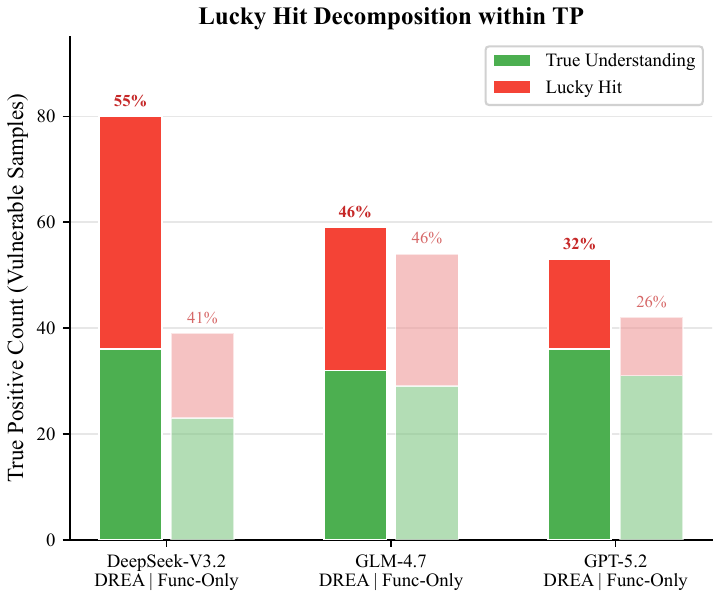}
\caption{Decomposition of true positives into correctly-reasoned detections and Lucky Hits (correct label, flawed reasoning) for each backbone under the DREA and Function-Only Baseline settings. Percentages above each bar indicate the Lucky Hit Rate.}
\Description{Stacked bar chart decomposing true positives into correctly-reasoned detections and Lucky Hits for three models under two methods, with Lucky Hit Rate percentages labeled.}
\label{fig:3}
\end{figure}

As Figure~\ref{fig:3} shows, a substantial fraction of true positives across both methods are Lucky Hits---correct labels supported by flawed reasoning. DeepSeek-V3.2 DREA reaches a Lucky Hit Rate of 55.0\%, GLM-4.7 DREA 45.8\%, and GPT-5.2 DREA 32.1\%.

The per-detection Reasoning Accuracy (RA) is lower for DREA than the baseline (e.g., DeepSeek-V3.2: 45\% vs.\ 59\%; GPT-5.2: 68\% vs.\ 74\%), but this reflects a composition effect: DREA detects substantially more true positives, and the additional detections tend to involve harder cases where correct reasoning is more difficult. In absolute terms, DREA yields more correctly-reasoned detections than the baseline in every configuration: DeepSeek-V3.2 36 vs.\ 23 (a 57\% increase), GPT-5.2 36 vs.\ 31, and GLM-4.7 32 vs.\ 29.

These results identify security reasoning quality as a shared bottleneck for current LLMs. Whether operating at the function level or with full repository access, all evaluated models produce a substantial proportion of correct predictions supported by flawed rationales. The model-wise differences further illustrate this point: GPT-5.2 achieves the lowest Lucky Hit Rate under both DREA and baseline settings, suggesting that reasoning quality is primarily determined by the backbone model's intrinsic security comprehension rather than by the amount of context provided.

\finding{5}{Reasoning correctness evaluation reveals that 26--55\% of true positives, for both DREA and the function-only baseline, are correct predictions supported by flawed rationales---identifying security reasoning quality as a shared bottleneck for current LLMs rather than a limitation specific to any detection paradigm.}

\section{Discussion}

\subsection{Practical Value of Hypothesis-Driven Exploration}

DREA demonstrates that hypothesis-driven repository exploration is a viable paradigm for vulnerability detection: authorization flaws invisible at the function level become detectable when DREA traces permission policies and compares sibling methods across the repository. Active, goal-directed context acquisition is the primary driver of improvement; the Planner--Explorer separation makes this economically practical and, as our ablation (Table~\ref{tab:ablation}) suggests, also shields the reasoning model from context overload. Not all vulnerabilities require repository-level evidence---cases with self-contained source-sink chains account for the baseline's 39--54\% recall---but DREA targets the complementary subset where cross-file evidence is essential.

The distinct model behavioral profiles (RQ3) have direct deployment implications. DREA's architecture supports swapping the Planner backbone without modifying the Explorer or tool infrastructure, allowing organizations to select aggressive (DeepSeek-V3.2) or conservative (GPT-5.2) operating points as needed.

\subsection{Security Reasoning as a Shared LLM Bottleneck}

The Lucky Hit phenomenon (RQ5) and the negative correlation between exploration intensity and correctness (RQ4) together point to a deeper implication: security reasoning quality, not information access, is the binding constraint on current LLM-based vulnerability detection. This bottleneck is shared across paradigms---it manifests similarly whether the model operates at the function level or with full repository access.

The asymmetry across vulnerability types sharpens this observation. Source-sink vulnerabilities with explicit data-flow structure can often be correctly reasoned about even with limited context, whereas absence-type flaws (e.g., missing authorization, inadequate output encoding) demand reasoning about what \emph{should} be present---a capability that current LLMs lack regardless of how much code they inspect. This aligns with Huang et al.'s~\cite{huangSemanticTrapFinetuned2026} finding that fine-tuned LLMs learn functional patterns rather than vulnerability root causes, and with Risse et al.'s~\cite{risseTopScoreWrong2025} demonstration that high benchmark scores can coexist with spurious correlations.

These observations suggest that future work should target security reasoning itself---through specialized training, reinforcement learning from security feedback~\cite{weyssowR2VulLearningReason2025}, specification-guided reasoning~\cite{zhuSpecificationguidedVulnerabilityDetection2025}, or architectures that enforce explicit evidence-grounding---rather than focusing solely on providing more context.

\subsection{Implications for Evaluation}
The Lucky Hit phenomenon has consequences for how the community evaluates LLM-based security tools. A model reporting 80\% recall appears to detect 80 out of 100 vulnerabilities; under reasoning correctness evaluation, only 36 of those 80 detections are accompanied by correct reasoning (RA\,=\,45\%). Paired metrics such as Pair-Correctness address the problem of inflated recall through over-reporting, but they likewise cannot distinguish genuine understanding from Lucky Hits. This concern echoes the practical findings of Steenhoek et al.~\cite{steenhoekClosingGapUser2025}, who report that professional developers find AI-powered vulnerability detection tools impractical due to high false positive rates and non-applicable fixes, and of Ami et al.~\cite{amiFalseNegativeThat2024}, who document industry practitioners' emphasis on false negatives as the more critical failure mode. We suggest that reasoning correctness evaluation provides a useful complement to standard binary metrics, whether through LLM-as-a-Judge, human evaluation of reasoning traces, or proxy measures such as CWE-type prediction accuracy.

\section{Related Work}

\subsection{LLM-Based Vulnerability Detection}

The dominant paradigm applies LLMs to isolated code snippets for binary vulnerability prediction, typically trained or evaluated on function-level datasets such as Devign~\cite{zhouDevignEffectiveVulnerability2019} and BigVul~\cite{fanCCCodeVulnerability2020}. Early work explored fine-tuning pre-trained code models such as CodeBERT~\cite{fengCodeBERTPreTrainedModel2020}, LineVul~\cite{fuLineVulTransformerbasedLinelevel2022}, and VulBERTa~\cite{hanifVulBERTaSimplifiedSource2022} on labeled vulnerability datasets, with later efforts incorporating multi-task learning~\cite{duGeneralizationEnhancedCodeVulnerability2024a,yangSecurityVulnerabilityDetection2024} and structured code representations~\cite{wenSCALEConstructingStructured2024}. More recently, instruction-tuned LLMs have been evaluated in zero-shot and few-shot settings: Ullah et al.~\cite{ullahLLMsCannotReliably2024} proposed the SecLLMHolmes framework and found that current LLMs exhibit non-deterministic outputs and unreliable reasoning, while Steenhoek et al.~\cite{steenhoekErrMachineVulnerability2025} evaluated 14 state-of-the-art models and reported balanced accuracy of only 50--55\%. Guo et al.~\cite{guoOutsideComfortZone2024} and Lin and Mohaisen~\cite{linLargeMammothComparative2025} confirm these limitations through systematic zero-shot evaluations. Li et al.~\cite{liEverythingYouWanted2025} provide a comprehensive evaluation that challenges prevailing beliefs about LLM unreliability by showing that context-rich settings improve performance substantially, and Liu et al.~\cite{liuVulDetectBenchEvaluatingDeep2024} introduce VulDetectBench to evaluate capabilities at multiple difficulty levels. Li et al.~\cite{liSVTrustEvalCEvaluatingStructure2025} further show that LLMs rely more on pattern matching than robust logical reasoning for vulnerability analysis. Huang et al.~\cite{huangSemanticTrapFinetuned2026} identify a ``semantic trap'' in fine-tuned LLMs, where models learn functional patterns rather than vulnerability root causes.

Chain-of-thought prompting has been shown to improve detection performance by reducing false positives and encouraging structured reasoning~\cite{nongChainofThoughtPromptingLarge2024,sunLLM4VulnUnifiedEvaluation2025}. Weyssow et al.~\cite{weyssowR2VulLearningReason2025} propose R2Vul, which uses reinforcement learning from AI feedback to teach models to generate security-aware explanations, while Zhu et al.~\cite{zhuSpecificationguidedVulnerabilityDetection2025} introduce specification-guided detection that extracts security specifications from historical vulnerabilities to improve both detection and reasoning. Several approaches attempt to bridge the context gap through static context augmentation: VulEval~\cite{wenVulEvalRepositoryLevelEvaluation2024} and ReposVul~\cite{wangReposVulRepositoryLevelHighQuality2024} append caller/callee functions, Yang et al.~\cite{yangContextEnhancedVulnerabilityDetection2025} extract multi-layer program context, Vul-RAG~\cite{duVulRAGEnhancingLLMbased2025} retrieves vulnerability knowledge via retrieval-augmented generation (RAG), LLM4Vuln~\cite{sunLLM4VulnUnifiedEvaluation2025} evaluates RAG-augmented vulnerability reports, and FocusVul~\cite{zhengLearningFocusContext2025} learns to select vulnerability-relevant context for efficient detection. LLMxCPG~\cite{lekssaysLLMxCPGContextAwareVulnerability2025} integrates Code Property Graphs with LLMs for cross-function analysis. Despite these advances, all approaches remain limited to goal-agnostic retrieval strategies that cannot adaptively explore a repository driven by the specific needs of a security investigation.

\subsection{LLM Agents for Software Engineering and Vulnerability Detection}

The agent paradigm equips LLMs with tools for code navigation and has achieved notable success in software engineering. SWE-Agent~\cite{yangSWEAgentAgentComputer2024} introduced an agent-computer interface for repository-level bug fixing. Agentless~\cite{xiaAgentlessDemystifyingLLMbased2024} showed that structured pipelines can achieve competitive results, and AutoCodeRover~\cite{zhangAutoCodeRoverAutonomousProgram2024} combines code search APIs with LLM reasoning for program repair.

Recent work has introduced agent-like mechanisms to vulnerability detection, but these approaches primarily use agents to retrieve statically extractable context. Yildiz et al.~\cite{yildizBenchmarkingLLMsLLMbased} equip a ReAct agent with \texttt{get\_callers}/\texttt{get\_callees} tools that extract pre-computed call graph information. Ahmed et al.~\cite{ahmedSecVulEvalBenchmarkingLLMs2025} construct a multi-agent pipeline where a context agent extracts function dependencies through static analysis. VulnLLM-R~\cite{nieVulnLLMRSpecializedReasoning2025} uses an agent scaffold to invoke CodeQL queries for context retrieval. MulVul~\cite{wuMulVulRetrievalaugmentedMultiagent2026} employs router and detector agents with retrieval tools to query a vulnerability knowledge base. AgenticSCR~\cite{charoenwetAgenticSCRAutonomousAgentic2026} targets immature vulnerabilities through agentic secure code review, and Xiong and Zhang~\cite{xiongSiftingNoiseComparative2026} compare agent frameworks for false positive filtering. Li et al.~\cite{liLLMbasedVulnerabilityDetection2026} present the first comprehensive empirical study of specialized LLM-based detectors at the project scale. Some approaches also draw on the broader idea of improving reasoning through multi-agent debate~\cite{irvingAISafetyDebate2018,duImprovingFactualityReasoning2023}. These systems augment LLMs with structured context access, but the agent's role remains bounded to searching within statically extracted information rather than autonomously navigating a live repository guided by evolving security hypotheses.

DREA differs in that the Explorer agent operates directly on the full project repository with general-purpose code navigation tools, and its exploration is directed by the Planner's security reasoning in real time. This hypothesis-driven exploration mirrors how a human auditor investigates a codebase and is fundamentally different from retrieving pre-extracted context. Additionally, our reasoning correctness evaluation reveals that 26--55\% of correct detections across all methods are supported by flawed rationales, identifying security reasoning quality as a shared bottleneck for current LLMs.

\section{Limitations}

Our benchmark covers 100 CVE pairs; while sufficient to observe the consistent differences reported in this paper, it may not capture long-tail vulnerability patterns. This scale reflects a deliberate quality-over-quantity trade-off, as each pair requires a recoverable repository snapshot, a single-function fix, and manual verification. We regard expanding the benchmark as an important direction for future work. The per-CWE analysis is based on small subgroups (3--13 pairs), limiting confidence in CWE-specific conclusions. Although our human validation study (Section~\ref{subsec:judge-reliability}) demonstrates strong alignment between the LLM judge and expert annotators (Cohen's $\kappa$ = 0.88--0.92), the human evaluation covered 50 cases; a larger-scale human study or multi-judge cross-validation could further strengthen confidence. The benchmark focuses exclusively on Python; DREA's design is language-agnostic, and extension to C/C++ is planned as future work. Public CVEs may overlap with LLM training data, but our pair-level protocol and the consistent improvements over same-backbone baselines indicate that gains arise from exploration rather than memorization. The Explorer agent (GLM-4.7-Flash) occasionally produces imperfect tool calls, and a more capable Explorer might improve retrieved context quality. Finally, benign labels denote the absence of the specific documented vulnerability rather than general security; if a model identifies a genuine but unrelated issue in a patched sample, our protocol counts it as a false positive. These limitations constrain the scope of our conclusions but do not undermine the core findings.

\section{Conclusion}
We present DREA, a hypothesis-driven framework that decouples security reasoning from repository exploration for repository-level vulnerability detection. Across three backbone models on RepoPairBench, DREA improves Pair-Correctness from 19--26\% to 30--42\% while offloading over 93\% of token consumption to a lightweight local Explorer, reducing estimated billable API cost by a factor of 16--48. Ablation experiments further support that the improvement arises from hypothesis-driven exploration rather than mere context volume. We additionally introduce a reasoning correctness evaluation and find that 26--55\% of correct detections, for both DREA and the function-only baseline, are supported by flawed rationales---identifying security reasoning quality as a shared bottleneck for current LLMs. Future work should pursue both directions: improving the agent architecture itself (e.g., more capable Explorers, cross-language support, hypothesis revision mechanisms) and strengthening the intrinsic security reasoning of LLMs through structured analytical training, reinforcement learning from security feedback, or architectures that enforce explicit evidence-grounding.

\begin{acks}
This work is supported in part by the Strategic Priority Research Program of Chinese Academy of Sciences (Grant No. XDB0690100), CAS Project for Young Scientists in Basic Research (Grant No. YSBR-118), and Beijing Natural Science Foundation (L253025).
\end{acks}

\bibliographystyle{ACM-Reference-Format}
\bibliography{references}

@misc{yangSWEAgentAgentComputer2024,
  title = {{SWE}-Agent: Agent-Computer Interfaces Enable Automated Software Engineering},
  author = {Yang, John and Jimenez, Carlos E. and Wettig, Alexander and Lieret, Kilian and Yao, Shunyu and Narasimhan, Karthik and Press, Ofir},
  year = 2024,
  month = may,
  number = {arXiv:2405.15793},
  eprint = {2405.15793},
  primaryclass = {cs},
  publisher = {arXiv},
  doi = {10.48550/arXiv.2405.15793},
  archiveprefix = {arXiv}
}

@misc{xiaAgentlessDemystifyingLLMbased2024,
  title = {Agentless: Demystifying {LLM}-Based Software Engineering Agents},
  author = {Xia, Chunqiu Steven and Deng, Yinlin and Dunn, Soren and Zhang, Lingming},
  year = 2024,
  month = jul,
  number = {arXiv:2407.01489},
  eprint = {2407.01489},
  primaryclass = {cs},
  publisher = {arXiv},
  doi = {10.48550/arXiv.2407.01489},
  archiveprefix = {arXiv}
}

@inproceedings{zhangAutoCodeRoverAutonomousProgram2024,
  title = {{AutoCodeRover}: Autonomous Program Improvement},
  author = {Zhang, Yuntong and Ruan, Haifeng and Fan, Zhiyu and Roychoudhury, Abhik},
  year = 2024,
  month = sep,
  booktitle = {Proceedings of the 33rd ACM SIGSOFT International Symposium on Software Testing and Analysis},
  pages = {1592--1604},
  publisher = {ACM},
  address = {Vienna, Austria},
  doi = {10.1145/3650212.3680384}
}

@misc{dingVulnerabilityDetectionCode2024,
  title = {Vulnerability Detection with Code Language Models: How Far Are We?},
  author = {Ding, Yangruibo and Fu, Yanjun and Ibrahim, Omniyyah and Sitawarin, Chawin and Chen, Xinyun and Alomair, Basel and Wagner, David and Ray, Baishakhi and Chen, Yizheng},
  year = 2024,
  month = mar,
  number = {arXiv:2403.18624},
  eprint = {2403.18624},
  primaryclass = {cs},
  publisher = {arXiv},
  doi = {10.48550/arXiv.2403.18624},
  archiveprefix = {arXiv}
}

@inproceedings{spadiniPyDrillerPythonFramework2018,
  title = {{PyDriller}: {Python} Framework for Mining Software Repositories},
  author = {Spadini, Davide and Aniche, Maur\'icio and Bacchelli, Alberto},
  year = 2018,
  month = oct,
  booktitle = {Proceedings of the 2018 26th ACM Joint Meeting on European Software Engineering Conference and Symposium on the Foundations of Software Engineering},
  pages = {908--911},
  publisher = {ACM},
  address = {Lake Buena Vista, FL, USA},
  doi = {10.1145/3236024.3264598}
}

@inproceedings{zhengJudgingLLMasaJudge2024,
  title = {Judging {LLM}-as-a-Judge with {MT}-Bench and Chatbot Arena},
  author = {Zheng, Lianmin and Chiang, Wei-Lin and Sheng, Ying and Zhuang, Siyuan and Wu, Zhanghao and Zhuang, Yonghao and Lin, Zi and Li, Zhuohan and Li, Dacheng and Xing, Eric P. and Zhang, Hao and Gonzalez, Joseph E. and Stoica, Ion},
  year = 2024,
  booktitle = {Advances in Neural Information Processing Systems},
  volume = {36},
  pages = {46595--46623},
  publisher = {Curran Associates, Inc.},
  address = {New Orleans, LA, USA}
}

@inproceedings{zhouDevignEffectiveVulnerability2019,
  title = {Devign: Effective Vulnerability Identification by Learning Comprehensive Program Semantics via Graph Neural Networks},
  author = {Zhou, Yaqin and Liu, Shangqing and Siow, Jingkai and Du, Xiaoning and Liu, Yang},
  year = 2019,
  booktitle = {Advances in Neural Information Processing Systems},
  volume = {32},
  pages = {10197--10207},
  publisher = {Curran Associates, Inc.},
  address = {Vancouver, BC, Canada}
}

@inproceedings{fanCCCodeVulnerability2020,
  title = {A {C/C++} Code Vulnerability Dataset with Code Changes and {CVE} Summaries},
  author = {Fan, Jiahao and Li, Yi and Wang, Shaohua and Nguyen, Tien N.},
  year = 2020,
  month = jun,
  booktitle = {Proceedings of the 17th International Conference on Mining Software Repositories},
  pages = {508--512},
  publisher = {ACM},
  address = {Seoul, Republic of Korea},
  doi = {10.1145/3379597.3387501}
}

@inproceedings{fengCodeBERTPreTrainedModel2020,
  title = {{CodeBERT}: A Pre-Trained Model for Programming and Natural Languages},
  author = {Feng, Zhangyin and Guo, Daya and Tang, Duyu and Duan, Nan and Feng, Xiaocheng and Gong, Ming and Shou, Linjun and Qin, Bing and Liu, Ting and Jiang, Daxin and Zhou, Ming},
  year = 2020,
  booktitle = {Findings of the Association for Computational Linguistics: EMNLP 2020},
  pages = {1536--1547},
  publisher = {Association for Computational Linguistics},
  address = {Online},
  doi = {10.18653/v1/2020.findings-emnlp.139}
}

@inproceedings{fuLineVulTransformerbasedLinelevel2022,
  title = {{LineVul}: A Transformer-Based Line-Level Vulnerability Prediction},
  author = {Fu, Michael and Tantithamthavorn, Chakkrit},
  year = 2022,
  month = may,
  booktitle = {Proceedings of the 19th International Conference on Mining Software Repositories},
  pages = {608--620},
  publisher = {ACM},
  address = {Pittsburgh, PA, USA},
  doi = {10.1145/3524842.3528452}
}

@inproceedings{hanifVulBERTaSimplifiedSource2022,
  title = {{VulBERTa}: Simplified Source Code Pre-Training for Vulnerability Detection},
  author = {Hanif, Hazim and Maffeis, Sergio},
  year = 2022,
  month = jul,
  booktitle = {2022 International Joint Conference on Neural Networks (IJCNN)},
  pages = {1--8},
  publisher = {IEEE},
  address = {Padua, Italy},
  doi = {10.1109/IJCNN55064.2022.9892280}
}

@inproceedings{ullahLLMsCannotReliably2024,
  title = {{LLMs} Cannot Reliably Identify and Reason About Security Vulnerabilities (Yet?): A Comprehensive Evaluation, Framework, and Benchmarks},
  author = {Ullah, Saad and Han, Mingji and Pujar, Saurabh and Pearce, Hammond and Coskun, Ayse and Stringhini, Gianluca},
  year = 2024,
  month = may,
  booktitle = {2024 IEEE Symposium on Security and Privacy (SP)},
  pages = {862--880},
  publisher = {IEEE},
  address = {San Francisco, CA, USA},
  doi = {10.1109/SP54263.2024.00210}
}

@misc{nongChainofThoughtPromptingLarge2024,
  title = {Chain-of-Thought Prompting of Large Language Models for Discovering and Fixing Software Vulnerabilities},
  author = {Nong, Yu and Aldeen, Mohammed and Cheng, Long and Hu, Hongxin and Chen, Feng and Cai, Haipeng},
  year = 2024,
  month = feb,
  number = {arXiv:2402.17230},
  eprint = {2402.17230},
  primaryclass = {cs},
  publisher = {arXiv},
  doi = {10.48550/arXiv.2402.17230},
  archiveprefix = {arXiv}
}

@misc{wangReposVulRepositoryLevelHighQuality2024,
  title = {{ReposVul}: A Repository-Level High-Quality Vulnerability Dataset},
  author = {Wang, Xinchen and Hu, Ruida and Gao, Cuiyun and Wen, Xin-Cheng and Chen, Yujia and Liao, Qing},
  year = 2024,
  month = jan,
  number = {arXiv:2401.13169},
  eprint = {2401.13169},
  primaryclass = {cs},
  publisher = {arXiv},
  doi = {10.48550/arXiv.2401.13169},
  archiveprefix = {arXiv}
}

@misc{irvingAISafetyDebate2018,
  title = {{AI} Safety via Debate},
  author = {Irving, Geoffrey and Christiano, Paul and Amodei, Dario},
  year = 2018,
  month = may,
  number = {arXiv:1805.00899},
  eprint = {1805.00899},
  primaryclass = {stat},
  publisher = {arXiv},
  doi = {10.48550/arXiv.1805.00899},
  archiveprefix = {arXiv}
}

@misc{duImprovingFactualityReasoning2023,
  title = {Improving Factuality and Reasoning in Language Models through Multiagent Debate},
  author = {Du, Yilun and Li, Shuang and Torralba, Antonio and Tenenbaum, Joshua B. and Mordatch, Igor},
  year = 2023,
  month = may,
  number = {arXiv:2305.14325},
  eprint = {2305.14325},
  primaryclass = {cs},
  publisher = {arXiv},
  doi = {10.48550/arXiv.2305.14325},
  archiveprefix = {arXiv}
}

@misc{ahmedSecVulEvalBenchmarkingLLMs2025,
  title = {{{SecVulEval}}: {{Benchmarking LLMs}} for {{Real-World C}}/{{C}}++ {{Vulnerability Detection}}},
  shorttitle = {{{SecVulEval}}},
  author = {Ahmed, Md Basim Uddin and Harzevili, Nima Shiri and Shin, Jiho and Pham, Hung Viet and Wang, Song},
  year = 2025,
  month = may,
  number = {arXiv:2505.19828},
  eprint = {2505.19828},
  primaryclass = {cs},
  publisher = {arXiv},
  doi = {10.48550/arXiv.2505.19828},
  urldate = {2026-03-14},
  archiveprefix = {arXiv},
}

@inproceedings{amiFalseNegativeThat2024,
  title = {"{{False}} Negative - That One Is Going to Kill You": {{Understanding Industry Perspectives}} of {{Static Analysis}} Based {{Security Testing}}},
  shorttitle = {"{{False}} Negative - That One Is Going to Kill You"},
  booktitle = {2024 {{IEEE Symposium}} on {{Security}} and {{Privacy}} ({{SP}})},
  author = {Ami, Amit Seal and Moran, Kevin and Poshyvanyk, Denys and Nadkarni, Adwait},
  year = 2024,
  month = may,
  pages = {3979--3997},
  publisher = {IEEE},
  address = {San Francisco, CA, USA},
  doi = {10.1109/SP54263.2024.00019},
  urldate = {2026-03-14},
  copyright = {https://doi.org/10.15223/policy-009},
  isbn = {979-8-3503-3130-1},
}

@misc{charoenwetAgenticSCRAutonomousAgentic2026,
  title = {{{AgenticSCR}}: {{An Autonomous Agentic Secure Code Review}} for {{Immature Vulnerabilities Detection}}},
  shorttitle = {{{AgenticSCR}}},
  author = {Charoenwet, Wachiraphan and Tantithamthavorn, Kla and Thongtanunam, Patanamon and Lin, Hong Yi and Jeong, Minwoo and Wu, Ming},
  year = 2026,
  month = jan,
  number = {arXiv:2601.19138},
  eprint = {2601.19138},
  primaryclass = {cs},
  publisher = {arXiv},
  doi = {10.48550/arXiv.2601.19138},
  urldate = {2026-03-14},
  archiveprefix = {arXiv},
  langid = {english},
}

@inproceedings{duGeneralizationEnhancedCodeVulnerability2024a,
  title = {Generalization-{{Enhanced Code Vulnerability Detection}} via {{Multi-Task Instruction Fine-Tuning}}},
  booktitle = {Findings of the {{Association}} for {{Computational Linguistics ACL}} 2024},
  author = {Du, Xiaohu and Wen, Ming and Zhu, Jiahao and Xie, Zifan and Ji, Bin and Liu, Huijun and Shi, Xuanhua and Jin, Hai},
  year = 2024,
  pages = {10507--10521},
  publisher = {Association for Computational Linguistics},
  address = {Bangkok, Thailand and virtual meeting},
  doi = {10.18653/v1/2024.findings-acl.625},
  urldate = {2026-03-14},
  langid = {english},
}

@misc{duVulRAGEnhancingLLMbased2025,
  title = {Vul-{{RAG}}: Enhancing {{LLM-based}} Vulnerability Detection via Knowledge-Level {{RAG}}},
  shorttitle = {Vul-{{RAG}}},
  author = {Du, Xueying and Zheng, Geng and Wang, Kaixin and Zou, Yi and Wang, Yujia and Deng, Wentai and Feng, Jiayi and Liu, Mingwei and Chen, Bihuan and Peng, Xin and Ma, Tao and Lou, Yiling},
  year = 2025,
  month = jun,
  number = {arXiv:2406.11147},
  eprint = {2406.11147},
  primaryclass = {cs},
  publisher = {arXiv},
  doi = {10.48550/arXiv.2406.11147},
  urldate = {2026-03-14},
  archiveprefix = {arXiv},
  langid = {english},
}

@misc{gaoMonoYourClean2025,
  title = {Mono: {{Is Your}} "{{Clean}}" {{Vulnerability Dataset Really Solvable}}? {{Exposing}} and {{Trapping Undecidable Patches}} and {{Beyond}}},
  shorttitle = {Mono},
  author = {Gao, Zeyu and Zhou, Junlin and Zhang, Bolun and He, Yi and Zhang, Chao and Cui, Yuxin and Wang, Hao},
  year = 2025,
  month = jun,
  number = {arXiv:2506.03651},
  eprint = {2506.03651},
  primaryclass = {cs},
  publisher = {arXiv},
  doi = {10.48550/arXiv.2506.03651},
  urldate = {2026-03-14},
  archiveprefix = {arXiv},
  langid = {english},
}

@misc{guoOutsideComfortZone2024,
  title = {Outside the Comfort Zone: Analysing {{LLM}} Capabilities in Software Vulnerability Detection},
  shorttitle = {Outside the Comfort Zone},
  author = {Guo, Yuejun and Patsakis, Constantinos and Hu, Qiang and Tang, Qiang and Casino, Fran},
  year = 2024,
  month = aug,
  number = {arXiv:2408.16400},
  eprint = {2408.16400},
  primaryclass = {cs},
  publisher = {arXiv},
  doi = {10.48550/arXiv.2408.16400},
  urldate = {2026-03-14},
  archiveprefix = {arXiv},
  langid = {english},
}

@misc{huangSemanticTrapFinetuned2026,
  title = {The Semantic Trap: Do Fine-Tuned {{LLMs}} Learn Vulnerability Root Cause or Just Functional Pattern?},
  shorttitle = {The Semantic Trap},
  author = {Huang, Feiyang and Sun, Yuqiang and Zhang, Fan and Yang, Ziqi and Liu, Han and Liu, Yang},
  year = 2026,
  month = feb,
  number = {arXiv:2601.22655},
  eprint = {2601.22655},
  primaryclass = {cs},
  publisher = {arXiv},
  doi = {10.48550/arXiv.2601.22655},
  urldate = {2026-03-14},
  archiveprefix = {arXiv},
  langid = {english},
}

@misc{kaniewskiSystematicLiteratureReview2025,
  title = {A Systematic Literature Review on Detecting Software Vulnerabilities with Large Language Models},
  author = {Kaniewski, Sabrina and Schmidt, Fabian and Enzweiler, Markus and Menth, Michael and Heer, Tobias},
  year = 2025,
  month = dec,
  number = {arXiv:2507.22659},
  eprint = {2507.22659},
  primaryclass = {cs},
  publisher = {arXiv},
  doi = {10.48550/arXiv.2507.22659},
  urldate = {2026-03-14},
  archiveprefix = {arXiv},
  langid = {english},
}

@misc{lekssaysLLMxCPGContextAwareVulnerability2025,
  title = {{{LLMxCPG}}: {{Context-Aware Vulnerability Detection Through Code Property Graph-Guided Large Language Models}}},
  shorttitle = {{{LLMxCPG}}},
  author = {Lekssays, Ahmed and Mouhcine, Hamza and Tran, Khang and Yu, Ting and Khalil, Issa},
  year = 2025,
  month = jul,
  number = {arXiv:2507.16585},
  eprint = {2507.16585},
  primaryclass = {cs},
  publisher = {arXiv},
  doi = {10.48550/arXiv.2507.16585},
  urldate = {2026-03-14},
  archiveprefix = {arXiv},
  langid = {english},
}

@misc{liCleanVulAutomaticFunctionLevel2025,
  title = {{{CleanVul}}: {{Automatic Function-Level Vulnerability Detection}} in {{Code Commits Using LLM Heuristics}}},
  shorttitle = {{{CleanVul}}},
  author = {Li, Yikun and Zhang, Ting and Widyasari, Ratnadira and Tun, Yan Naing and Nguyen, Huu Hung and Bui, Tan and Irsan, Ivana Clairine and Cheng, Yiran and Lan, Xiang and Ang, Han Wei and Liauw, Frank and Weyssow, Martin and Kang, Hong Jin and Ouh, Eng Lieh and Shar, Lwin Khin and Lo, David},
  year = 2025,
  month = sep,
  number = {arXiv:2411.17274},
  eprint = {2411.17274},
  primaryclass = {cs},
  publisher = {arXiv},
  doi = {10.48550/arXiv.2411.17274},
  urldate = {2026-03-14},
  archiveprefix = {arXiv},
  langid = {english},
}

@misc{liEverythingYouWanted2025,
  title = {Everything You Wanted to Know about {{LLM-based}} Vulnerability Detection but Were Afraid to Ask},
  author = {Li, Yue and Li, Xiao and Wu, Hao and Xu, Minghui and Zhang, Yue and Cheng, Xiuzhen and Xu, Fengyuan and Zhong, Sheng},
  year = 2025,
  month = apr,
  number = {arXiv:2504.13474},
  eprint = {2504.13474},
  primaryclass = {cs},
  publisher = {arXiv},
  doi = {10.48550/arXiv.2504.13474},
  urldate = {2026-03-14},
  archiveprefix = {arXiv},
  langid = {english},
}

@misc{liLLMbasedVulnerabilityDetection2026,
  title = {{{LLM-based}} Vulnerability Detection at Project Scale: An Empirical Study},
  shorttitle = {{{LLM-based}} Vulnerability Detection at Project Scale},
  author = {Li, Fengjie and Jiang, Jiajun and Chen, Dongchi and Xiong, Yingfei},
  year = 2026,
  month = jan,
  number = {arXiv:2601.19239},
  eprint = {2601.19239},
  primaryclass = {cs},
  publisher = {arXiv},
  doi = {10.48550/arXiv.2601.19239},
  urldate = {2026-03-14},
  archiveprefix = {arXiv},
  langid = {english},
}

@inproceedings{linLargeMammothComparative2025,
  title = {From Large to Mammoth: A Comparative Evaluation of Large Language Models in Zero-Shot Vulnerability Detection},
  shorttitle = {From Large to Mammoth},
  booktitle = {Proceedings 2025 {{Network}} and {{Distributed System Security Symposium}}},
  author = {Lin, Jie and Mohaisen, David},
  year = 2025,
  publisher = {Internet Society},
  address = {San Diego, CA, USA},
  doi = {10.14722/ndss.2025.241491},
  urldate = {2026-03-14},
  isbn = {979-8-9894372-8-3},
  numpages = {18},
  langid = {english},
}

@misc{liSVTrustEvalCEvaluatingStructure2025,
  title = {{{SV-TrustEval-C}}: {{Evaluating Structure}} and {{Semantic Reasoning}} in {{Large Language Models}} for {{Source Code Vulnerability Analysis}}},
  shorttitle = {{{SV-TrustEval-C}}},
  author = {Li, Yansong and Branco, Paula and Hoole, Alexander M. and Marwah, Manish and Koduvely, Hari Manassery and Jourdan, Guy-Vincent and Jou, Stephan},
  year = 2025,
  month = may,
  number = {arXiv:2505.20630},
  eprint = {2505.20630},
  primaryclass = {cs},
  publisher = {arXiv},
  doi = {10.48550/arXiv.2505.20630},
  urldate = {2026-03-14},
  archiveprefix = {arXiv},
}

@misc{liuVulDetectBenchEvaluatingDeep2024,
  title = {{{VulDetectBench}}: Evaluating the Deep Capability of Vulnerability Detection with Large Language Models},
  shorttitle = {{{VulDetectBench}}},
  author = {Liu, Yu and Gao, Lang and Yang, Mingxin and Xie, Yu and Chen, Ping and Zhang, Xiaojin and Chen, Wei},
  year = 2024,
  month = aug,
  number = {arXiv:2406.07595},
  eprint = {2406.07595},
  primaryclass = {cs},
  publisher = {arXiv},
  doi = {10.48550/arXiv.2406.07595},
  urldate = {2026-03-14},
  archiveprefix = {arXiv},
  langid = {english},
}

@misc{nieVulnLLMRSpecializedReasoning2025,
  title = {{{VulnLLM-R}}: Specialized Reasoning {{LLM}} with Agent Scaffold for Vulnerability Detection},
  shorttitle = {{{VulnLLM-R}}},
  author = {Nie, Yuzhou and Li, Hongwei and Guo, Chengquan and Jiang, Ruizhe and Wang, Zhun and Li, Bo and Song, Dawn and Guo, Wenbo},
  year = 2025,
  month = dec,
  number = {arXiv:2512.07533},
  eprint = {2512.07533},
  primaryclass = {cs},
  publisher = {arXiv},
  doi = {10.48550/arXiv.2512.07533},
  urldate = {2026-03-14},
  archiveprefix = {arXiv},
  langid = {english},
}

@misc{risseTopScoreWrong2025,
  title = {Top {{Score}} on the {{Wrong Exam}}: {{On Benchmarking}} in {{Machine Learning}} for {{Vulnerability Detection}}},
  shorttitle = {Top {{Score}} on the {{Wrong Exam}}},
  author = {Risse, Niklas and Liu, Jing and B{\"o}hme, Marcel},
  year = 2025,
  month = apr,
  number = {arXiv:2408.12986},
  eprint = {2408.12986},
  primaryclass = {cs},
  publisher = {arXiv},
  doi = {10.48550/arXiv.2408.12986},
  urldate = {2026-03-14},
  archiveprefix = {arXiv},
  langid = {english},
}

@misc{shengLLMsSoftwareSecurity2025,
  title = {{{LLMs}} in Software Security: A Survey of Vulnerability Detection Techniques and Insights},
  shorttitle = {{{LLMs}} in Software Security},
  author = {Sheng, Ze and Chen, Zhicheng and Gu, Shuning and Huang, Heqing and Gu, Guofei and Huang, Jeff},
  year = 2025,
  month = feb,
  number = {arXiv:2502.07049},
  eprint = {2502.07049},
  primaryclass = {cs},
  publisher = {arXiv},
  doi = {10.48550/arXiv.2502.07049},
  urldate = {2026-03-14},
  archiveprefix = {arXiv},
  langid = {english},
}

@misc{steenhoekClosingGapUser2025,
  title = {Closing the Gap: A User Study on the Real-World Usefulness of {{AI-powered}} Vulnerability Detection \& Repair in the {{IDE}}},
  shorttitle = {Closing the Gap},
  author = {Steenhoek, Benjamin and Sivaraman, Kalpathy and Gonzalez, Renata Saldivar and Mohylevskyy, Yevhen and Moghaddam, Roshanak Zilouchian and Le, Wei},
  year = 2025,
  month = apr,
  number = {arXiv:2412.14306},
  eprint = {2412.14306},
  primaryclass = {cs},
  publisher = {arXiv},
  doi = {10.48550/arXiv.2412.14306},
  urldate = {2026-03-14},
  archiveprefix = {arXiv},
  langid = {english},
}

@misc{steenhoekErrMachineVulnerability2025,
  title = {To Err Is Machine: Vulnerability Detection Challenges {{LLM}} Reasoning},
  shorttitle = {To Err Is Machine},
  author = {Steenhoek, Benjamin and Rahman, Md Mahbubur and Roy, Monoshi Kumar and Alam, Mirza Sanjida and Tong, Hengbo and Das, Swarna and Barr, Earl T. and Le, Wei},
  year = 2025,
  month = jan,
  number = {arXiv:2403.17218},
  eprint = {2403.17218},
  primaryclass = {cs},
  publisher = {arXiv},
  doi = {10.48550/arXiv.2403.17218},
  urldate = {2026-03-14},
  archiveprefix = {arXiv},
  langid = {english},
}

@misc{sunLLM4VulnUnifiedEvaluation2025,
  title = {{{LLM4Vuln}}: A Unified Evaluation Framework for Decoupling and Enhancing {{LLMs}}' Vulnerability Reasoning},
  shorttitle = {{{LLM4Vuln}}},
  author = {Sun, Yuqiang and Wu, Daoyuan and Xue, Yue and Liu, Han and Ma, Wei and Zhang, Lyuye and Liu, Yang and Li, Yingjiu},
  year = 2025,
  month = jun,
  number = {arXiv:2401.16185},
  eprint = {2401.16185},
  primaryclass = {cs},
  publisher = {arXiv},
  doi = {10.48550/arXiv.2401.16185},
  urldate = {2026-03-14},
  archiveprefix = {arXiv},
  langid = {english},
}

@inproceedings{wenSCALEConstructingStructured2024,
  title = {{{SCALE}}: Constructing Structured Natural Language Comment Trees for Software Vulnerability Detection},
  shorttitle = {Scale},
  booktitle = {Proceedings of the 33rd {{ACM SIGSOFT International Symposium}} on {{Software Testing}} and {{Analysis}}},
  author = {Wen, Xin-Cheng and Gao, Cuiyun and Gao, Shuzheng and Xiao, Yang and Lyu, Michael R.},
  year = 2024,
  month = sep,
  pages = {235--247},
  publisher = {ACM},
  address = {Vienna Austria},
  doi = {10.1145/3650212.3652124},
  urldate = {2026-03-14},
  isbn = {979-8-4007-0612-7},
  langid = {english},
}

@misc{wenVulEvalRepositoryLevelEvaluation2024,
  title = {{{VulEval}}: {{Towards Repository-Level Evaluation}} of {{Software Vulnerability Detection}}},
  shorttitle = {{{VulEval}}},
  author = {Wen, Xin-Cheng and Wang, Xinchen and Chen, Yujia and Hu, Ruida and Lo, David and Gao, Cuiyun},
  year = 2024,
  month = apr,
  number = {arXiv:2404.15596},
  eprint = {2404.15596},
  primaryclass = {cs},
  publisher = {arXiv},
  doi = {10.48550/arXiv.2404.15596},
  urldate = {2026-03-14},
  archiveprefix = {arXiv},
  langid = {english},
}

@misc{weyssowR2VulLearningReason2025,
  title = {{{R2Vul}}: {{Learning}} to {{Reason}} about {{Software Vulnerabilities}} with {{Reinforcement Learning}} and {{Structured Reasoning Distillation}}},
  shorttitle = {{{R2Vul}}},
  author = {Weyssow, Martin and Yang, Chengran and Chen, Junkai and Widyasari, Ratnadira and Zhang, Ting and Huang, Huihui and Nguyen, Huu Hung and Tun, Yan Naing and Bui, Tan and Li, Yikun and Wei, Ang Han and Liauw, Frank and Ouh, Eng Lieh and Shar, Lwin Khin and Lo, David},
  year = 2025,
  month = aug,
  number = {arXiv:2504.04699},
  eprint = {2504.04699},
  primaryclass = {cs},
  publisher = {arXiv},
  doi = {10.48550/arXiv.2504.04699},
  urldate = {2026-03-14},
  archiveprefix = {arXiv},
  langid = {english},
}

@misc{wuMulVulRetrievalaugmentedMultiagent2026,
  title = {{{MulVul}}: Retrieval-Augmented Multi-Agent Code Vulnerability Detection via Cross-Model Prompt Evolution},
  shorttitle = {{{MulVul}}},
  author = {Wu, Zihan and Xu, Jie and Peng, Yun and Chong, Chun Yong and Jia, Xiaohua},
  year = 2026,
  month = jan,
  number = {arXiv:2601.18847},
  eprint = {2601.18847},
  primaryclass = {cs},
  publisher = {arXiv},
  doi = {10.48550/arXiv.2601.18847},
  urldate = {2026-03-14},
  archiveprefix = {arXiv},
  langid = {english},
}

@misc{xiongSiftingNoiseComparative2026,
  title = {Sifting the Noise: A Comparative Study of {{LLM}} Agents in Vulnerability False Positive Filtering},
  shorttitle = {Sifting the Noise},
  author = {Xiong, Yunpeng and Zhang, Ting},
  year = 2026,
  month = jan,
  number = {arXiv:2601.22952},
  eprint = {2601.22952},
  primaryclass = {cs},
  publisher = {arXiv},
  doi = {10.48550/arXiv.2601.22952},
  urldate = {2026-03-14},
  archiveprefix = {arXiv},
  langid = {english},
}

@misc{yangContextEnhancedVulnerabilityDetection2025,
  title = {Context-{{Enhanced Vulnerability Detection Based}} on {{Large Language Model}}},
  author = {Yang, Yixin and Xu, Bowen and Gao, Xiang and Sun, Hailong},
  year = 2025,
  month = apr,
  number = {arXiv:2504.16877},
  eprint = {2504.16877},
  primaryclass = {cs},
  publisher = {arXiv},
  doi = {10.48550/arXiv.2504.16877},
  urldate = {2026-03-14},
  archiveprefix = {arXiv},
  langid = {english},
}

@misc{yangSecurityVulnerabilityDetection2024,
  title = {Security Vulnerability Detection with Multitask Self-Instructed Fine-Tuning of Large Language Models},
  author = {Yang, Aidan Z. H. and Tian, Haoye and Ye, He and Martins, Ruben and Goues, Claire Le},
  year = 2024,
  month = jun,
  number = {arXiv:2406.05892},
  eprint = {2406.05892},
  primaryclass = {cs},
  publisher = {arXiv},
  doi = {10.48550/arXiv.2406.05892},
  urldate = {2026-03-14},
  archiveprefix = {arXiv},
  langid = {english},
}

@inproceedings{yildizBenchmarkingLLMsLLMbased,
  title = {Benchmarking {{LLMs}} and {{LLM-based Agents}} in {{Practical Vulnerability Detection}} for {{Code Repositories}}},
  author = {Yildiz, Alperen and Teo, Sin G and Lou, Yiling and Feng, Yebo and Wang, Chong and Divakaran, Dinil Mon},
  year = 2025,
  booktitle = {Proceedings of the 63rd Annual Meeting of the Association for Computational Linguistics (Volume 1: Long Papers)},
  pages = {30848--30865},
  publisher = {Association for Computational Linguistics},
  address = {Vienna, Austria},
  doi = {10.18653/v1/2025.acl-long.1490},
  langid = {english},
}

@misc{zhengLearningFocusContext2025,
  title = {Learning to Focus: Context Extraction for Efficient Code Vulnerability Detection with Language Models},
  shorttitle = {Learning to Focus},
  author = {Zheng, Xinran and Qian, Xingzhi and Zhou, Huichi and Yang, Shuo and He, Yiling and Jana, Suman and Cavallaro, Lorenzo},
  year = 2025,
  month = jul,
  number = {arXiv:2505.17460},
  eprint = {2505.17460},
  primaryclass = {cs},
  publisher = {arXiv},
  doi = {10.48550/arXiv.2505.17460},
  urldate = {2026-03-14},
  archiveprefix = {arXiv},
  langid = {english},
}

@misc{zhouLargeLanguageModel2024,
  title = {Large Language Model for Vulnerability Detection and Repair: Literature Review and the Road Ahead},
  shorttitle = {Large Language Model for Vulnerability Detection and Repair},
  author = {Zhou, Xin and Cao, Sicong and Sun, Xiaobing and Lo, David},
  year = 2024,
  month = oct,
  number = {arXiv:2404.02525},
  eprint = {2404.02525},
  primaryclass = {cs},
  publisher = {arXiv},
  doi = {10.48550/arXiv.2404.02525},
  urldate = {2026-03-14},
  archiveprefix = {arXiv},
  langid = {english},
}

@misc{zhuSpecificationguidedVulnerabilityDetection2025,
  title = {VulInstruct: Teaching LLMs Root-Cause Reasoning for Vulnerability Detection via Security Specifications},
  author = {Zhu, Hao and Li, Jia and Gao, Cuiyun and Qian, Jiaru and Dong, Yihong and Liu, Huanyu and Wang, Lecheng and Wang, Ziliang and Hu, Xiaolong and Li, Ge},
  year = 2026,
  number = {arXiv:2511.04014},
  eprint = {2511.04014},
  archiveprefix = {arXiv},
  primaryclass = {cs.SE},
  publisher = {arXiv},
  doi = {10.48550/arXiv.2511.04014},
  url = {https://arxiv.org/abs/2511.04014},
  urldate = {2026-03-14},
  langid = {english},
}

\end{document}